\documentclass[pdflatex,sn-mathphys-num]{sn-jnl}% Math and Physical Sciences Numbered Reference Style 
\usepackage{graphicx}%
\usepackage{multirow}%
\usepackage{amsmath,amssymb,amsfonts}%
\usepackage{amsthm}%
\usepackage{mathrsfs}%
\usepackage[title]{appendix}%
\usepackage{xcolor}%
\usepackage{textcomp}%
\usepackage{manyfoot}%
\usepackage{booktabs}%
\usepackage{algorithm}%
\usepackage{algorithmicx}%
\usepackage{algpseudocode}%
\usepackage{listings}%
\usepackage{mdframed}
\usepackage{multicol}
\theoremstyle{thmstyleone}%
\theoremstyle{thmstyletwo}%

\theoremstyle{thmstylethree}%

\begin{document}

\title{Self-Certification of High-Risk AI Systems: The Example of AI-based Facial Emotion Recognition}


\author*[1]{\fnm{Gregor} \sur{Autischer}}\email{gregor.autischer@student.tugraz.at}

\author[2]{\fnm{Kerstin} \sur{Waxnegger}}\email{kwaxnegger@know-center.at}

\author*[2,3]{\fnm{Dominik} \sur{Kowald}}\email{dkowald@know-center.at}

\affil[1]{\orgname{Graz University of Technology}, \orgaddress{\city{Graz}, \country{Austria}}}

\affil[2]{\orgname{Know Center Research GmbH}, \orgaddress{\city{Graz}, \country{Austria}}}

\affil[3]{\orgname{University of Graz}, \orgaddress{\city{Graz}, \country{Austria}}}


\abstract{The European Union's Artificial Intelligence Act establishes comprehensive requirements for high-risk AI systems, yet the harmonized standards necessary for demonstrating compliance remain not fully developed. In this paper, we investigate the practical application of the Fraunhofer AI assessment catalogue as a certification framework through a complete self-certification cycle of an AI-based facial emotion recognition system. Beginning with a baseline model that has deficiencies, including inadequate demographic representation and prediction uncertainty, we document an enhancement process guided by AI certification requirements. The enhanced system achieves higher accuracy with improved reliability metrics and comprehensive fairness across demographic groups. We focused our assessment on two of the six Fraunhofer catalogue dimensions, reliability and fairness, the enhanced system successfully satisfies the certification criteria for these examined dimensions. We find that the certification framework provides value as a proactive development tool, driving concrete technical improvements and generating documentation naturally through integration into the development process. However, fundamental gaps separate structured self-certification from legal compliance: harmonized European standards are not fully available, and AI assessment frameworks and catalogues cannot substitute for them on their own. These findings establish the Fraunhofer AI assessment catalogue as a valuable preparatory tool that complements rather than replaces formal compliance requirements at this time.}


\maketitle
\section{Introduction}\label{sec1}
Artificial Intelligence (AI) has evolved from research labs into a ubiquitous technology that touches virtually every sector of modern society. Modern AI systems now operate across many domains, from healthcare and financial services to semi-autonomous vehicles and personalized recommender systems \citep{kowald2024establishing,mullner2023differential}. Virtual assistants shape interactions, algorithmic decision-making systems influence credit approval and hiring processes and machine learning models help optimize energy consumption \citep{kasinidou_artificial_2024}. This rapid proliferation of AI applications across critical societal functions has fundamentally transformed how individuals, organizations and governments interact with this technology \citep{polak_exploring_2024}.

However, this widespread adoption introduces challenges that extend beyond technical performance considerations. Concerns regarding fairness, transparency, accountability and potential societal impacts have emerged as central issues \citep{baum_fear_2023}. Systems that influence fundamental rights, such as biometric identification, emotion recognition and automated decision-making in employment or law enforcement, raise important questions about bias, discrimination and the protection of individual autonomy. The complexity of modern AI systems, combined with their often opaque decision-making processes, complicates efforts to ensure these technologies operate safely and ethically \citep{cheong_transparency_2024}. These challenges have created the need for comprehensive regulatory frameworks that can address the risks AI systems present while still allowing for innovation \citep{ayub_regulating_2024}.

The European Union has responded to these challenges by creating the AI Act, a landmark regulatory framework \citep{european-union_regulation_2024}. The legislation represents the most comprehensive attempt worldwide to regulate AI systems through a risk-based approach, with requirements that scale with the potential harm a system might cause \citep{mueck_etsi_2022}. The AI Act establishes clear obligations for providers, deployers and other actors across the AI value chain, with particular emphasis on high-risk applications \citep{pimentel_why_2024}. Other regulatory initiatives have also emerged globally \citep{al-maamari_between_2025, isaperl_state_2025}. These developments change the landscape for AI development and deployment, requiring systems to demonstrate compliance with specific technical, ethical and governance requirements.

As a result, some AI systems must now undergo examination to verify compliance with the regulation. AI certification differs from traditional software auditing, as AI systems have characteristics that require a different auditing approach \citep{winter_trusted_2021, isaca2024aiauditors}. Organizations developing or deploying AI systems face the practical challenge of demonstrating that their technologies meet the requirements set by policymakers. However, the certification of AI systems remains a relatively new and evolving domain, with limited practical guidance on how theoretical frameworks translate into a compliant system \citep{brogle2025context}. To address this gap, we investigate the application of the Fraunhofer AI Assessment Catalogue as a self-certification framework for a facial emotion recognition system, evaluating its effectiveness and if it is possible to achieve AI Act compliance.

\subsection{Our Previous Research}
\label{sec:previous_work}
This paper builds directly on our prior research. In \citep{autischer_ai_2025, autischer_practical_2025}, we attempted a sample certification of the RIOT Art installation with EmoPy, an open-source facial emotion recognition system, using the Fraunhofer AI Assessment Catalogue \citep{poretschkin_ai_2023, thoughtworks_thoughtworksartsriot_2019, perez_emopy_2018}. In this work, we build upon that foundation and extend our investigation substantially. We describe our approach in detail in Section~\ref{sec3}.

\subsection{Research Questions}
\label{sec:research_questions}
Our previous papers identified gaps between the AI system and requirements from the certification process. This paper continues that investigation by conducting a complete self-certification cycle, in which we actively modify the system to address identified shortcomings. We pursue three primary research questions:

\textbf{RQ1:} What practical insights emerge from conducting a complete self-certification cycle of an AI system using the Fraunhofer AI Assessment Catalogue?

\textbf{RQ2:} What does technical testing reveal that documentation-only certification can miss?

\textbf{RQ3:} To what extent does this self-certification address the requirements established by the AI Act?

These questions examine the entire certification process. We contribute a complete self-certification cycle of an AI system using the Fraunhofer AI Assessment Catalogue, demonstrate how technical testing reveals reliability issues that documentation alone could miss and clarify the current gap between such frameworks and formal regulatory compliance.

\section{Foundations and Background}
\label{sec2}
In this section, we present the necessary prerequisites for this paper. We examine four key elements: the European AI Act as the regulatory framework, the developing norms and standards that operationalize its requirements, the Fraunhofer AI Assessment Catalogue we use for this self-certification and an overview of the facial emotion recognition system that we attempt to self-certify.

\subsection{The European AI Act, Standards and Certification}
\label{sec:ai_act_norms}
The European Union's Artificial Intelligence Act (AI Act) represents the most comprehensive regulatory initiative in the field of AI, establishing a risk-based framework that categorizes AI systems according to their potential to cause harm \citep{european-union_regulation_2024}. The legislation entered into force on August 1, 2024, creating legally binding requirements for providers, deployers and other actors across the AI value chain \citep{mueck_etsi_2022}. The AI Act addresses concerns about AI fairness, transparency, accountability and potential societal impacts \citep{pimentel_why_2024, baum_fear_2023}.
The AI Act employs a risk-based categorization system that imposes different requirements depending on the potential harm a system might cause. High-risk AI systems, including those used for biometric identification, emotion recognition, or law enforcement, face the strictest requirements. Facial emotion recognition systems, the focus of this research, fall under the AI Act as high-risk applications according to Article 6(2) in conjunction with Annex III, point 1(c). These systems must demonstrate compliance with the technical requirements set out in Articles 9–15, including risk management (Art. 9), data governance (Art. 10), technical documentation (Art. 11), record-keeping (Art. 12), transparency (Art. 13), human oversight (Art. 14) and accuracy, robustness and cybersecurity (Art. 15) \citep{european-union_regulation_2024}.

\subsubsection{Certification}
To prove adherence to these regulatory requirements, organizations must conduct auditing and certification processes that verify AI systems meet certain standards. However, certification cannot occur in a regulatory vacuum. It requires concrete, measurable norms and standards against which evaluators can assess systems \citep{werry_eu_2024}. The AI Act explicitly anticipates this need, mandating the development of harmonized European standards that provide presumption of conformity with the regulation's requirements \citep{kilian_european_2025}.

\subsubsection{Standards for the AI Act}
The European Commission tasked CEN (European Committee for Standardization) and CENELEC (European Committee for Electrotechnical Standardization) with developing these critical standards through their Joint Technical Committee 21 (JTC 21), established specifically for AI standardization \citep{cen-cenelec_jtc_21_artificial_2025, cen-cenelec_jtc_21_european_2024}. This committee is developing harmonized standards supporting the AI Act implementation. The committee's work encompasses AI trustworthiness frameworks, risk management standards, quality management systems and conformity assessment procedures designed to translate abstract regulatory requirements into concrete technical specifications.

These norms matter greatly because certification processes depend on evaluating systems against established standards. When a system achieves certification to a harmonized European standard, it gains presumption of conformity with the corresponding AI Act requirements once authorities publish these standards in the Official Journal of the European Union (OJEU) \citep{leyden_standards_2025, european-union_regulation_2024}. This mechanism provides the crucial link between abstract legal obligations and demonstrable technical compliance. Without appropriate standards, organizations lack clear guidance on what specific measures satisfy regulatory requirements and certification bodies lack objective criteria for evaluation \citep{obrien_role_2024}.

However, the standards development process faces significant delays. While the AI Act entered into force in August 2024, the comprehensive set of harmonized standards required for full implementation remain incomplete. The European Commission formally revised the original deadline and current delays indicate that work on the standards will continue into 2026 \citep{kroet_eu_2025}. This delay creates uncertainty for organizations seeking to demonstrate compliance, as the specific technical specifications they must meet remain partially undefined.

\subsubsection{ISO/IEC 42001}
Some existing international standards provide partial guidance. ISO/IEC 42001, the International AI Management Standard published in December 2023, offers a framework for establishing, implementing, maintaining and continually improving an AI management system within organizations. Companies and certification bodies have begun offering certification programs \citep{iso_isoiec_2023}. However, this standard represents a management system standard rather than a technical specification for individual AI systems. Therefore, achieving ISO 42001 certification does not automatically mean compliance with the AI Act \citep{erleblebici_iso_2025}.

\subsection{Fraunhofer AI Assessment Catalogue}
\label{sec:fraunhofer_catalogue}
At Fraunhofer the AI Assessment Catalogue was created with explicit consideration of the emerging AI Act, but it was completed before the regulation's finalization. Therefore, it cannot prove compliance on its own \citep{poretschkin_ai_2023}. The catalogue provides a structured approach for evaluating AI applications. Unlike ISO/IEC 42001's organizational focus, the Fraunhofer Catalogue offers detailed criteria to analyze technical characteristics and operational properties. It is structured like a detailed questionnaire that assessors can use to walk through a system step by step to analyze its properties.
The catalogue organizes its assessment approach around six dimensions of AI trustworthiness: fairness, autonomy and control, transparency, reliability, safety and security and data protection. Each dimension contains multiple specific criteria that allow for detailed evaluation. The framework guides assessors to gather technical data, documentation and operational data that address each relevant criterion. This evidence then forms the basis for arguments demonstrating how the system satisfies these trustworthiness requirements.
This approach addresses the unique challenges AI systems present compared to traditional software, where decision-making patterns emerge through learning from data rather than explicit programming. The framework's comprehensiveness makes it valuable for self-certification efforts, helping organizations identify specific areas requiring improvement to potentially meet anticipated regulatory standards.

\subsection{The Facial Emotion Recognition System}
\label{sec:fer_system}
\begin{figure}
	\begin{center}
		\includegraphics[width=.65\textwidth]{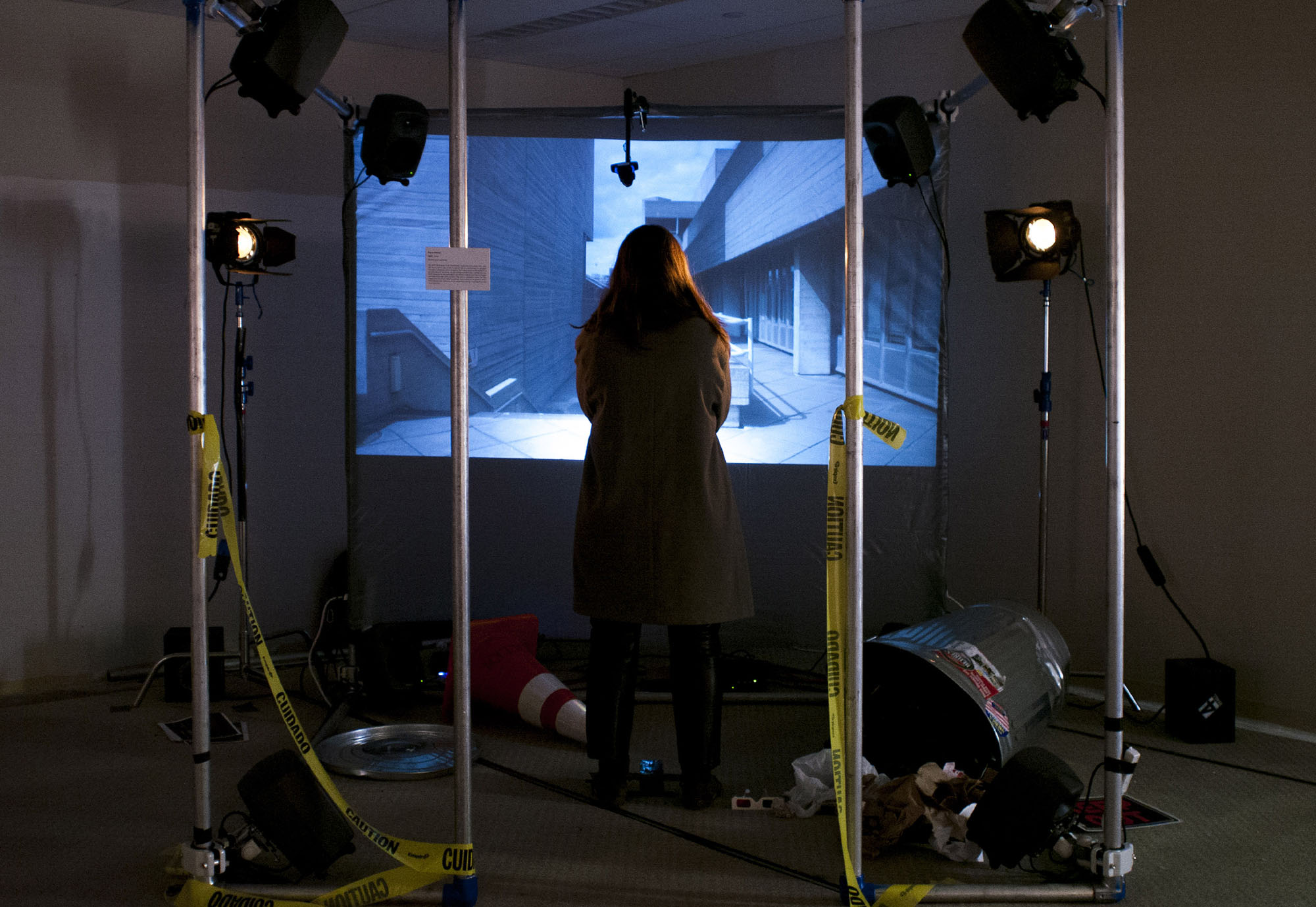}
	\end{center}
	\caption[RIOT Installation in New York 2018]{\textbf{RIOT Installation in New York 2018}. This image shows the RIOT art installation with a participant experiencing the interactive film. \citep{perez_emopy_2018}}
	\label{fig:riot_installation_background}
\end{figure}

We use a facial emotion recognition system for this self-certification, selected specifically for its relevance to the AI Act. Emotion recognition systems fall within the high-risk category of the regulation, subjecting them to comprehensive regulatory requirements \citep{european-union_regulation_2024}. The complete AI system is the RIOT interactive art installation, which employs facial emotion recognition to create a responsive narrative experience. As shown in Figure~\ref{fig:riot_installation_background}, participants stand before a screen displaying an interactive film while a mounted webcam captures their facial expressions \citep{thoughtworks_riot_2018, palmer_riot_2016}. The AI component analyzes these expressions and adapts the film's progression based on detected emotional states, creating a dynamic viewing experience where audience emotions change the movie's narrative.

Originally, the RIOT installation utilized the EmoPy framework for emotion recognition \citep{angelica_perez_thoughtworksartsemopy_2021}. For this paper, we reconstructed the original EmoPy Convolutional Neural Network model in PyTorch, creating the EmoTorch implementation, and, for the sake of reproducibility in AI \citep{semmelrock2025reproducibility}, we publish it via GitHub \citep{autischer_gregor-autischeremotorch_paper_2025}. This reconstruction serves two purposes: it establishes a baseline model that replicates the original system's architecture and behavior and it provides a foundation for implementing improvements. Building from this baseline, we developed an improved version that addresses the shortcomings identified during the previous certification attempt \citep{autischer_practical_2025}. The improved model represents the current EmoTorch implementation that underwent the complete self-certification cycle we document in this paper.

\subsubsection{Compilation of System Resources}

To facilitate the certification process, we provide here a complete overview of all available information and documentation for the system. We treat all items listed here as documentation resources for the certification process.

\paragraph{RIOT Installation}
\begin{itemize}
    \item GitHub repository of the RIOT art installation \citep{thoughtworks_thoughtworksartsriot_2019}
    \item Article on the RIOT art installation \citep{thoughtworks_riot_2018}
    \item TED Talk on the RIOT art installation \citep{ted_residency_karen_2018}
    \item Article describing different art installations including the RIOT art installation \citep{thoughtworks_thoughtworks_2018}
    \item Article on Karen Palmer, the artist behind the RIOT art installation \citep{thoughtworks_karen_2017}
    \item Short description of the RIOT art installation by Karen Palmer \citep{palmer_riot_2016}
    \item Video showcasing and describing the RIOT art installation \citep{karen_palmer_riot_2017}
\end{itemize}

\paragraph{Original EmoPy Framework}
\begin{itemize}
    \item GitHub repository of the EmoPy framework \citep{angelica_perez_thoughtworksartsemopy_2021}
    \item Article describing the EmoPy framework and technical decisions \citep{perez_emopy_2018}
    \item Article describing the architecture of the emotion recognition model \citep{perez_recognizing_2018}
    \item Python documentation of the EmoPy framework \citep{emopy_development_team_welcome_2017}
\end{itemize}

\paragraph{EmoTorch Model and Datasets}
\begin{itemize}
    \item GitHub repository of EmoTorch containing both the baseline rebuilt model and the enhanced model with complete documentation and technical information \citep{autischer_gregor-autischeremotorch_paper_2025}
    \item GitHub repository of EmoTorch in a wrapper for easy emotion recognition \citep{autischer_gregor-autischeremotorch_2025}
    \item GitHub repository of the adapted FairFace project used to create fairness labels for the training data \citep{autischer_gregor-autischeremotorch_run_fairface_2025}
    \item GitHub repository of the FER+ dataset \citep{microsoft_microsoftferplus_2023}
    \item Extended Cohn-Kanade dataset \citep{cohn_resources_2024}
    \item Real-world Affective Faces (RAF) dataset \citep{noauthor_real-world_nodate}
\end{itemize}

\section{Methodology}
\label{sec3}
In this section, we describe our methodological approach for the self-certification cycle of the facial emotion recognition system. Our methodology builds directly on the previous papers that identified certification gaps, but extends that investigation by actively addressing shortcomings through iterative improvements. This approach transforms the certification from passive evaluation into an active development process, revealing practical insights about making AI systems certifiable.

\subsection{Foundation in Previous Papers}
\label{sec:methodology_foundation}

The methodological foundation for this paper originates from the sample certification we described in the prior papers \citep{autischer_practical_2025, autischer_ai_2025}. In that initial investigation, we applied the Fraunhofer AI Assessment Catalogue to the RIOT project's emotion recognition system, which originally used the EmoPy framework. Our primary objective in that certification attempt was not to validate the system's quality, but rather to explore and understand the certification process itself: where assessment criteria prove challenging to meet and what documentation certification needs. That sample certification also revealed several areas where the system failed to meet the requirements.

However, that preliminary effort encountered a significant limitation: the EmoPy framework was no longer actively used or developed, which prevented us from addressing the identified issues. Without active development or the ability to modify the system, we could only document deficiencies without exploring how to fix them. Consequently, the certification process reduced to a binary pass-or-fail assessment rather than an iterative improvement process.

In this work, we address this limitation by rebuilding the facial emotion recognition system from the ground up and conducting a complete self-certification cycle with the Fraunhofer AI Assessment Catalogue. This approach enables iterative system improvements guided by certification requirements. By improving the system with certification as an explicit design objective, we investigate the practical challenges that emerge when attempting to make real systems certifiable.

\subsection{System Reconstruction}
\label{sec:system_reconstruction}

To enable continued certification work that could actively address identified shortcomings, we needed a system with an adaptable codebase and all the information on the datasets that are used. This necessitated reconstructing the system from its foundations.

We centered the reconstruction process on faithfully rebuilding EmoPy's convolutional neural network architecture using PyTorch, creating the EmoTorch implementation \citep{autischer_gregor-autischeremotorch_paper_2025}. This rebuild aimed to replicate the original model's behavior as closely as possible, ensuring that the baseline EmoTorch model served as a functional equivalent to the original EmoPy system. We drew on available documentation including confusion matrices, performance metrics and architectural descriptions from the original EmoPy documentation \citep{perez_emopy_2018, perez_recognizing_2018, angelica_perez_thoughtworksartsemopy_2021}. By carefully matching these documented characteristics, the baseline model maintained continuity with the original system while enabling the development of enhancements.

This decision represents a pragmatic compromise. Treating the baseline EmoTorch model as equivalent to the original EmoPy system allows our certification investigation to continue where the previous papers concluded, maintaining focus on the process of addressing certification requirements rather than starting the assessment from the beginning. The reconstruction does not compromise our research objectives, as the investigation centers on understanding the certification process. These process-oriented questions remain independent of the exact details of the underlying implementation, as long as the functional behavior remains equivalent. We acknowledge that the rebuilt model may not behave identically to the original in all edge cases, but since our research focuses on the certification process rather than the specific system, this limitation does not affect our findings.

\subsection{Self-Certification Process}
\label{sec:self_certification_process}
With the baseline EmoTorch model established, we proceeded through a systematic self-certification cycle designed to identify deficiencies, implement targeted improvements and evaluate resulting enhancements. This methodology transforms certification from retrospective evaluation into prospective development, where assessment findings directly lead to system improvements.

We began the self-certification with the baseline EmoTorch model, using shortcomings identified in the previous certification \citep{autischer_practical_2025} as the starting point. Following the Fraunhofer AI Assessment Catalogue's structured approach, we focused on two key dimensions: reliability and fairness. We determined this focus through the protection requirement analysis, which identified these dimensions as carrying medium risk for this application context. We did not investigate other dimensions that also carried medium risk due to time constraints. Moreover, reliability and fairness represent key dimensions for AI trustworthiness and since our primary interest lies in understanding the certification process rather than achieving exhaustive coverage, this focused scope remains appropriate.

The first certification attempt showed that the baseline model lacked adequate evaluation data for demographic fairness assessment. Training and testing datasets contained insufficient data on racial diversity, age ranges and other demographic characteristics necessary for comprehensive fairness evaluation (e.g., \citep{scher2023modelling}. We then used adapted training and evaluation datasets that provided the required information. The technical testing during the improvement process revealed some additional shortcomings in documentation, testing procedures and performance consistency. We addressed all these identified issues during the improvement process. We provide all modifications, testing results, documentation updates and the finished enhanced model in the EmoTorch repository \citep{autischer_gregor-autischeremotorch_paper_2025}.

We concluded the self-certification with a complete reassessment using the Fraunhofer AI Assessment Catalogue. In this final evaluation, we applied the same systematic criteria we used in the initial assessment, but now to the improved EmoTorch model. We completed the catalogue questionnaire for the dimensions under examination (reliability and fairness), documenting how the improved system addresses each relevant criterion.

\section{Initial Certification Assessment \& System Enhancement}
\label{sec4}

Building upon our prior papers where we applied the Fraunhofer certification catalogue to an initial baseline model \citep{autischer_practical_2025}, this section describes the enhancements we made to the facial emotion recognition system to address the identified certification deficiencies.

\subsection{Baseline Model Certification Assessment}
\label{sec:baseline_assessment}

The baseline emotion recognition system represents the initial implementation we used for this certification. In this section, we describe in detail the model architecture, training configuration, dataset properties and performance outcomes that then lead to the enhanced model.

\subsubsection{Baseline Model Architecture and Configuration}
\label{subsec:baseline_architecture}

The baseline system uses a Convolutional Neural Network (CNN) architecture designed for facial emotion classification. The network consists of two convolutional blocks, each containing 10 filters with a 4$\times$4 kernel. Feature extraction follows a straightforward approach where convolutional outputs undergo flattening before passing through a fully connected layer. The architecture applies ReLU activation at the final layer and has approximately 1,700 trainable parameters. For training, we use Cosine Proximity Loss, optimised via RMSprop with a learning rate of 0.0005. The learning rate decreases when validation performance plateaus. We use early stopping with patience of 3 epochs.
\subsubsection{Baseline Dataset Characteristics}
\label{subsec:baseline_dataset}

For baseline model training, we utilise two facial expression datasets: FER-2013 and the Extended Cohn-Kanade Dataset (CK+) \citep{microsoft_microsoftferplus_2023, cohn_resources_2024}. Together, they provide 20,940 images partitioned into 16,962 training samples and 3,620 validation samples. The classification task addresses four emotion categories: anger, fear, calm and surprise. We apply 28 data augmentation transformation types per image to expand the effective training set.

\begin{figure}[htbp]
    \centering
    \includegraphics[width=\linewidth]{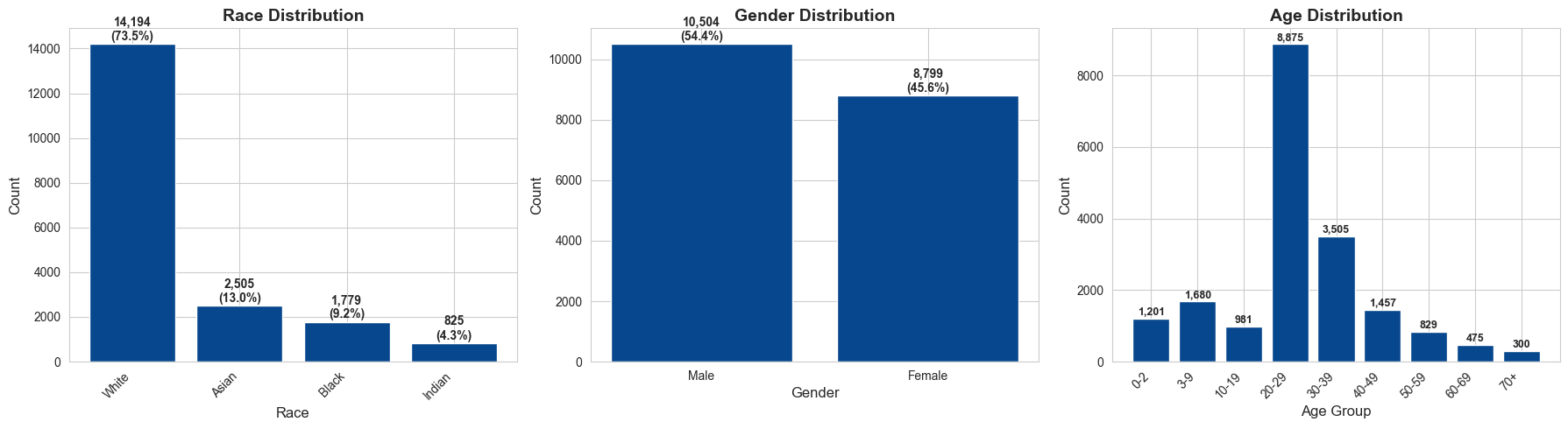}
    \caption{Baseline dataset composition showing data distribution across race, gender and age.}
    \label{fig:baseline_dataset}
\end{figure}

Several limitations characterise this dataset. Demographic metadata was entirely unavailable in the initial implementation, making any fairness analysis across gender, race, or age dimensions impossible. This absence of demographic information represented a fundamental barrier to certification and was one of the primary areas where the baseline model failed to satisfy certification requirements. To begin addressing this deficiency and start the development of an improved model, we examined these factors for the initial dataset. As no annotations for the required features were available, we used the FairFace library to create those features reliably. The resulting distributions can be seen in Figure~\ref{fig:baseline_dataset}. Racial representation showed substantial skew toward Caucasians, gender distribution appeared relatively balanced and age representation concentrated in the middle-age groups.

\subsubsection{Baseline Performance Results}
\label{subsec:baseline_performance}

\begin{figure}[htbp]
    \centering
    \includegraphics[width=\linewidth]{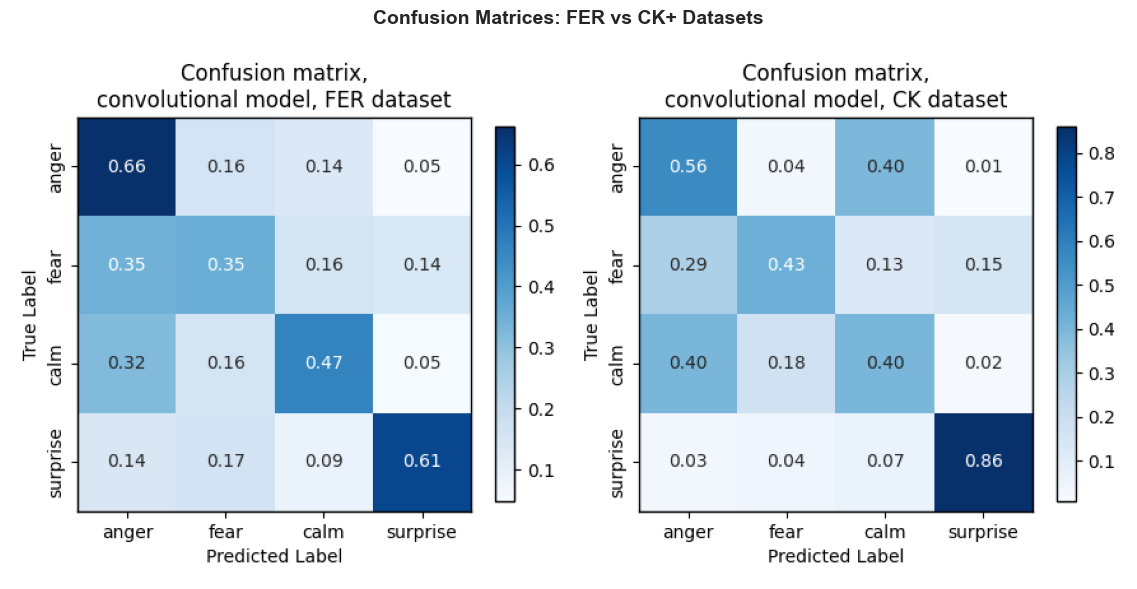}
    \caption{Confusion matrices of baseline models.}
    \label{fig:baseline_architecture}
\end{figure}

Our evaluation of the baseline model revealed substantial performance limitations. Combined validation accuracy reaches 53.88\%, with FER validation at 51.29\% and CK+ validation at 60.80\%. We primarily documented performance through confusion matrices for the two datasets. Figure~\ref{fig:baseline_architecture} shows these matrices for the rebuilt baseline model.

These metrics obscured a critical underlying issue that only emerged during technical testing. Our testing revealed that the model exhibited severe prediction uncertainty, a characteristic not apparent from the accuracy documentation alone. The model only achieved confidence levels of about 27\% for a 4-class prediction problem, with an entropy value of 1.38 indicating high uncertainty in classification decisions. This finding represents a significant reliability concern: while the model produced nominally correct outputs at acceptable rates, the low confidence score suggests that this model is not dependable for classification.

\subsection{Identified Deficiencies}
\label{sec:certification_deficiencies}

The reliance on basic accuracy metrics during the initial certification process proved insufficient for identifying critical underlying problems. While the reported accuracy of approximately 54\% provided a general performance indicator, this metric masked prediction uncertainty that only emerged through deeper investigation. The model demonstrated confidence levels of merely 27\% even for correct predictions, a finding not apparent from the original accuracy reporting. This shows the importance of comprehensive documentation with metrics capable of revealing multiple failure modes rather than relying solely on accuracy measurements.

A system that produces correct classifications with minimal certainty cannot be considered dependable for operational contexts where consistent, trustworthy predictions are essential. Furthermore, our demographic analysis of the training data revealed problematic dataset composition. The imbalances in age representation, racial distribution and the complete absence of augmentation data for people wearing head coverings collectively indicate that data quality investigation is essential.
These findings demonstrate that comprehensive evaluation is necessary for meaningful certification assessment. We describe the subsequent enhancements below, which address each of the identified shortcomings.

\subsection{System Enhancement}
\label{sec:enhancement_strategy}

To address the identified deficiencies, we improved the model architecture, training and dataset. We set four primary objectives for the model enhancement. First, we must address certification gaps identified in the initial assessment. Second, we must substantially improve overall model accuracy. Third, we must enable comprehensive fairness evaluation through appropriate dataset annotation. Fourth, we must enhance robustness and generalization capabilities to support reliable operation across varied conditions.

\subsubsection{Architecture Enhancements}
\label{subsec:architecture_enhancements}

The enhanced system also uses a CNN architecture, but with several modifications to address the baseline model's limitations. We increased the network depth to three convolutional blocks with progressive filter counts of 32, 64 and 128. We reduced kernel dimensions to 3$\times$3 with padding of 1, preserving spatial dimensions throughout the convolutional stages. Batch normalisation layers follow each convolutional block, which stabilises training and enables higher learning rates. We apply progressive dropout rates of 0.2, 0.3 and 0.4 across successive blocks to mitigate overfitting risk. Global Average Pooling replaces the flatten operation. We expanded the fully connected portion to include an intermediate layer, going from 128 to 256 to 4 units. The complete architecture now comprises over 300,000 trainable parameters.

\subsubsection{Training Improvements}
\label{subsec:training_improvements}

For training, we replaced Cosine Proximity Loss with weighted CrossEntropyLoss, addressing class imbalance by assigning higher weights to underrepresented emotion categories. We use the AdamW optimiser with learning rate 0.001 and weight decay 0.01, which leads to improved convergence. For learning rate scheduling, we now use CosineAnnealingWarmRestarts. We increased early stopping patience to 10 epochs, permitting longer training trajectories before termination. We implement class balancing through WeightedRandomSampler, which oversamples minority classes during batch construction. Gradient clipping with maximum norm 1.0 ensures training stability by preventing gradient explosion.

\subsubsection{Dataset Enhancements}
\label{subsec:dataset_enhancements}

\begin{figure}[htbp]
    \centering
    \includegraphics[width=\linewidth]{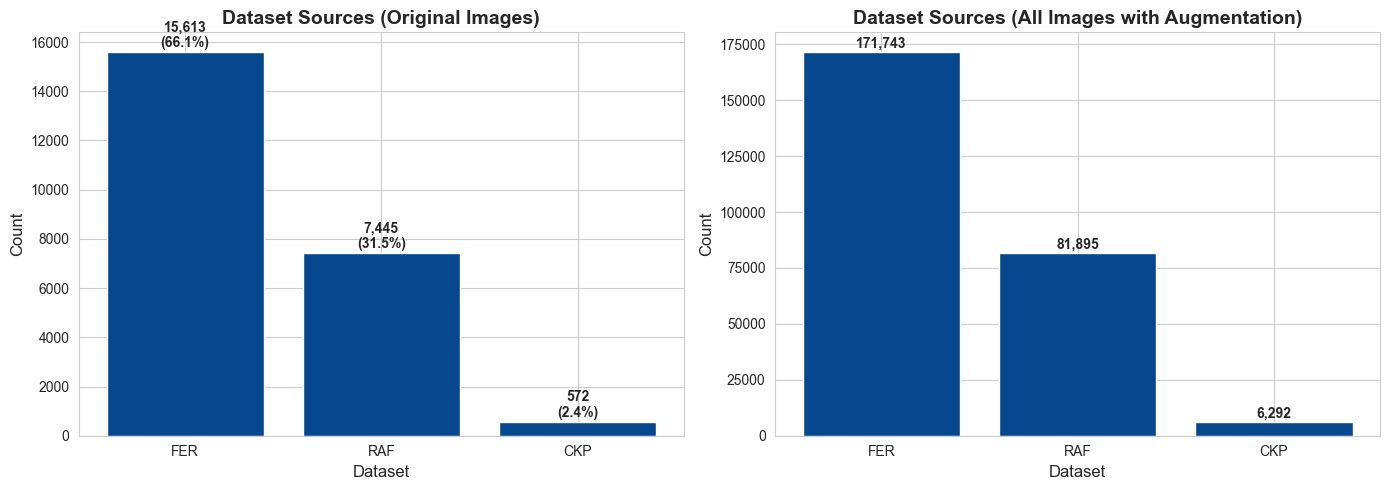}
    \caption{The new dataset that was used for training and evaluating the enhanced model.}
    \label{fig:enhanced_architecture}
\end{figure}

We expanded the dataset by adding RAF-DB to the existing FER-2013 and CK+ datasets \citep{microsoft_microsoftferplus_2023, cohn_resources_2024, li2019reliable}. The enhanced dataset comprises 23,630 original images, expanding to 259,930 samples with augmentation. Figure~\ref{fig:enhanced_architecture} shows the distribution of the data sources. We shifted the augmentation strategy from 28 generic transformations to more focused augmentation types designed to simulate realistic deployment conditions. These include rotation, dark lighting simulation, high contrast adjustment, light noise injection, blur application, head covering simulation through rectangular and diagonal occlusions, forehead bar occlusion representing headbands and hair strand simulation.

We gave explicit attention to class balance by reducing "Calm" class samples to 50\% of the original volume. Most critically, we added demographic metadata annotations for each image, enabling better fairness analysis. Annotations include gender categories (male, female, unsure), race categories (Caucasian, African-American, Asian) and age group categories (0-3, 4-19, 20-39, 40-69, 70+).

\subsection{The Enhanced Model}
\label{sec:enhanced_evaluation}

Our evaluation of the enhanced model shows many improvements across performance, fairness and robustness. We describe the results here. This enhanced model then underwent the certification process with the Fraunhofer Catalogue.

\subsubsection{Overall Performance Results}
\label{subsec:overall_performance}

\begin{figure}[htbp]
    \centering
    \includegraphics[width=\linewidth]{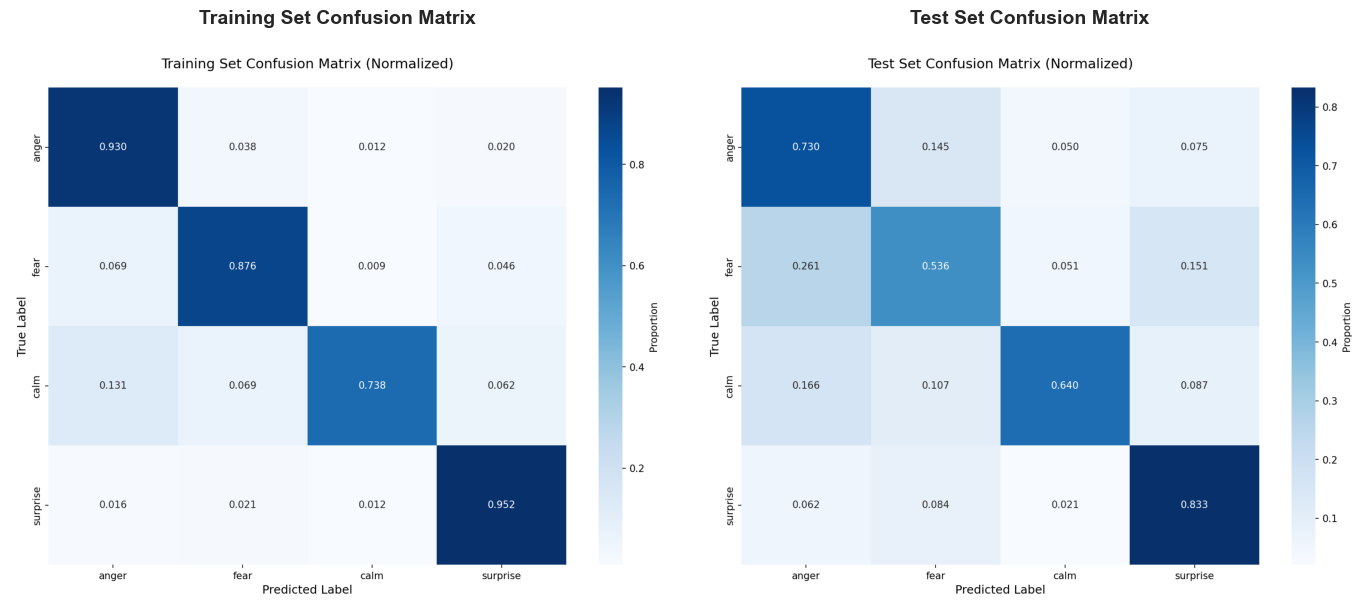}
    \caption{The confusion matrices of the enhanced model.}
    \label{fig:enhanced_dataset}
\end{figure}

The enhanced model achieves a test accuracy of 68.19\%, representing an improvement over the baseline model. Macro F1-score reaches 0.676, indicating balanced performance across emotion categories. Training accuracy of 78.75\% yields a training-test gap of approximately 10.5\%, suggesting appropriate model capacity without severe overfitting. Figure~\ref{fig:enhanced_dataset} shows the confusion matrices for the enhanced model, which demonstrate clear improvements over the baseline model.

\paragraph{Demographic Fairness Analysis}
\label{subsec:fairness_analysis}

\begin{figure}[htbp]
    \centering
    \includegraphics[width=\linewidth]{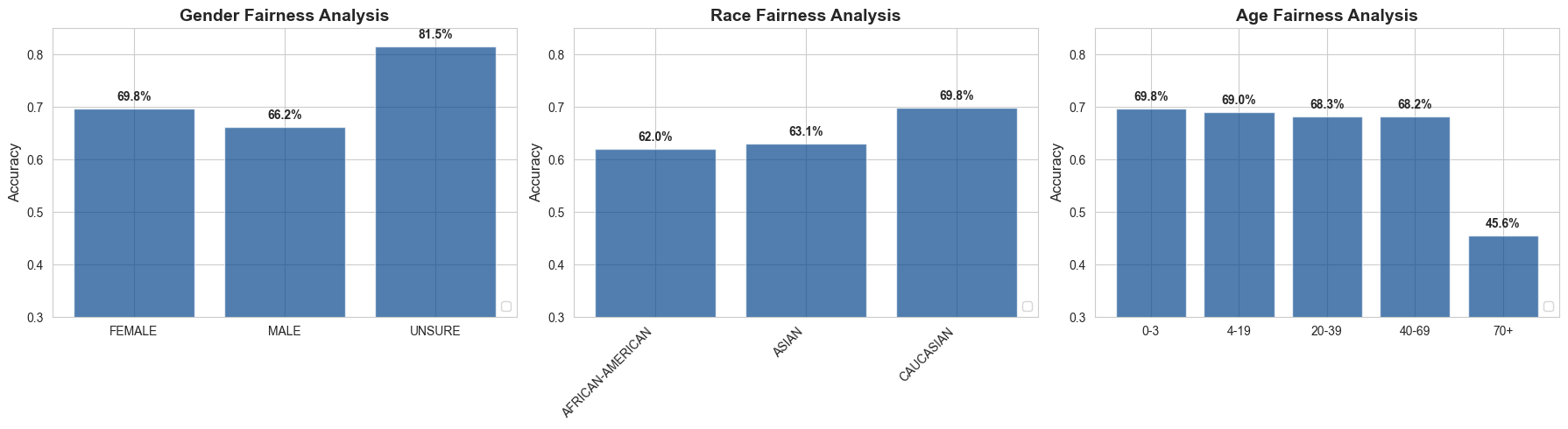}
    \caption{Demographic fairness analysis showing accuracy distribution across gender, race and age categories.}
    \label{fig:fairness_analysis}
\end{figure}

The availability of demographic metadata enables fairness evaluation across gender, race and age, addressing a critical gap we identified in the baseline assessment. Figure~\ref{fig:fairness_analysis} provides a fairness overview of the model's performance across these dimensions.
Our gender evaluation reveals accuracy of 69.78\% for female and 66.22\% for male faces, yielding an accuracy gap of 3.56 percentage points.
Our race analysis demonstrates accuracy of nearly 70\% for Caucasian, 67\% for Asian and 62\% for African-American people. The maximum accuracy gap of 7.85 percentage points between the highest and lowest performing groups remains within an acceptable range.

Our age group analysis reveals strong performance exceeding 68\% accuracy for people aged 0-69. However, the 70+ age group achieves only 45.61\% accuracy. This finding indicates that elderly people require special consideration, representing a known limitation for the real-world system. We did not address this gap, as we could not find additional training data for the 70+ age group and elderly individuals are not the primary target audience for the RIOT art installation. For other deployment contexts, this limitation would require resolution before certification.

\paragraph{Augmentation Robustness}
\label{subsec:robustness_results}

\begin{figure}[htbp]
    \centering
    \includegraphics[width=0.6\textwidth]{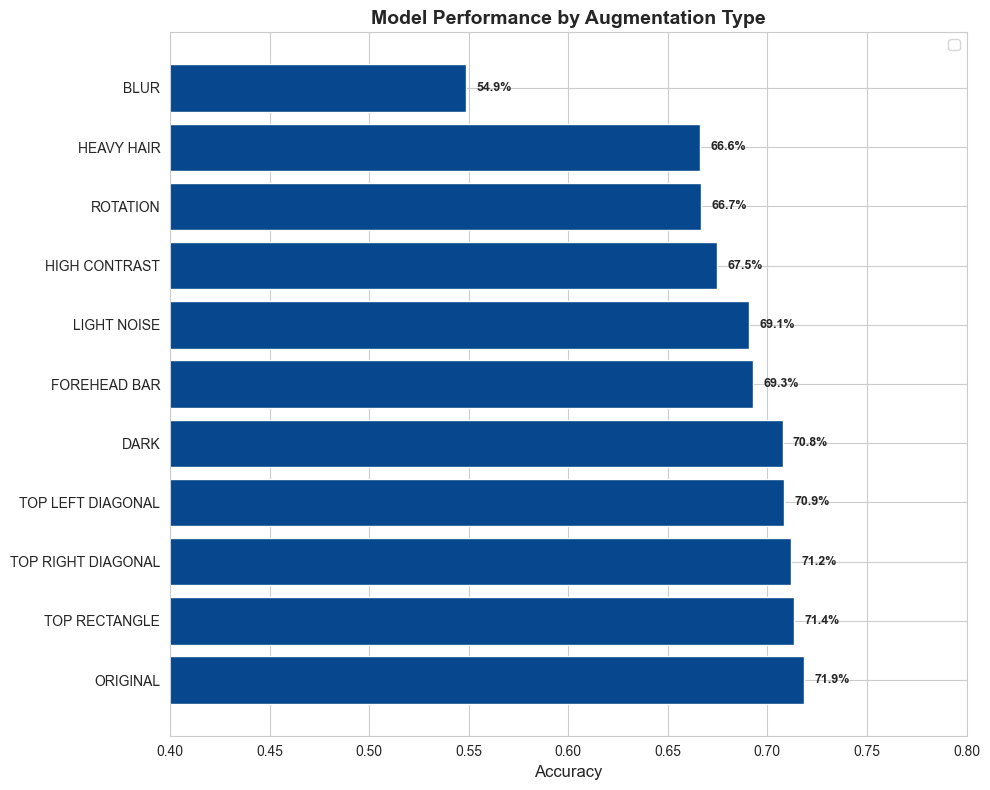}
    \caption{The performance of the enhanced model for different data augmentations.}
    \label{fig:training_curves}
\end{figure}

Our augmentation evaluation shows that the enhanced model maintains accuracy across all augmentation types. The blur augmentation shows the weakest performance (54.89\%). All other augmentations, including occlusion-based augmentations (simulating hijabs, caps and hair), maintain performance above 66\%, validating the system's performance for these scenarios. This performance indicates resilience to real-world variations, including different lighting conditions, partial facial occlusions from head coverings and hair and image quality degradation through noise. Only very blurry images lead to more severe degradation of prediction accuracy, which is acceptable as this can be well controlled in the deployment. Figure~\ref{fig:training_curves} illustrates these results.

\section{Self-Certification Assessment}
\label{sec5}

In the previous section, we detailed the enhancements to the baseline emotion recognition system, addressing the identified shortcomings. This section presents the results of applying the Fraunhofer AI Assessment Catalogue to the enhanced system.
Following our development of the enhanced model architecture and the comprehensive fairness and reliability evaluation, we repeated the complete certification process using the Fraunhofer Catalogue. We summarize the key findings and certification outcomes here.

\subsection{Certification Scope and Process}
\label{sec:cert_scope}

Our self-certification assessment focused on the two dimensions we already assessed in the previous paper: fairness and reliability. We applied each criterion from the Fraunhofer Catalogue systematically, documenting compliance and identifying any possible remaining gaps. We note that this constitutes a self-certification, where we evaluate our own system. This inherently differs from independent third-party certification, which would provide additional validation.

\subsection{Certification Findings}
\label{sec:cert_findings}

The certification assessment revealed that the enhancements we implemented have successfully addressed the previously identified shortcomings.
For the reliability dimension, the enhanced model meets all established targets. Test accuracy of 68.19\% exceeds the 60\% minimum threshold. Prediction confidence improved substantially from 27\% in the baseline model to 78.7\% and entropy decreased from 1.38 to 0.53, indicating more decisive classifications. Our expanded evaluation methodology, incorporating metrics such as entropy and confidence scoring alongside traditional accuracy metrics, provides the comprehensive assessment required by the certification framework.

For the fairness dimension, demographic gaps remain within acceptable bounds across all tested categories. The gender accuracy and racial gaps fall within acceptable limits, as we already discussed. The documented demographic distribution of the training data, combined with per-group accuracy documentation, satisfies the catalogue requirements.

Based on this assessment, the enhanced emotion recognition system satisfies the certification requirements of the Fraunhofer AI Certification Catalogue for the fairness and reliability dimensions.

\section{Regulatory Analysis and Compliance Gap}
\label{sec6}

This section examines how the self-certification we conducted in this paper relates to the AI Act. As we established in Section~\ref{sec:ai_act_norms}, the AI Act mandates a conformity assessment for high-risk systems before market placement, with harmonized standards providing presumption of conformity \citep{european-union_regulation_2024}.

\subsection{The Standards and Certification Gap}
\label{sec:standards-gap}

Two fundamental gaps separate the self-certification we conducted here from AI Act compliance. First, the harmonized European standards required for conformity assessment remain under development by CEN-CENELEC JTC 21, with delays beyond the original 2025 target \citep{kilian_european_2025, eu_artificial_intelligence_act_standard_2025}. Without published harmonized standards cited in the Official Journal of the European Union, systems cannot benefit from presumption of conformity under Article 40 \citep{european-union_regulation_2024, wong_role_2024}.

Second, the Fraunhofer AI Assessment Catalogue, while valuable for structured evaluation, is not a harmonized standard and does not fulfill mandatory conformity assessment procedures \citep{poretschkin_ai_2023}. The conformity assessment under the AI Act requires specific procedures, resulting in the EU Declaration of Conformity \citep{european-union_regulation_2024, euaiact_key_2024}. The catalogue assessment we conducted in this paper does not constitute this legally required conformity process \citep{holistic_conformity_2023}.

As we noted in Section~\ref{sec:ai_act_norms}, ISO/IEC 42001 certification does not constitute AI Act compliance, as it represents a management system standard rather than a technical specification for individual AI systems \citep{iso_isoiec_2023, chatard_eu_2024, vanta_how_2025}. Organizations must recognize the critical distinction between organizational management standards and product conformity assessment.

\subsection{Value for System Development}
\label{sec:development-value}

Despite these gaps, the Fraunhofer Catalogue-based self-certification provided substantial value for internal system development. The structured assessment across reliability and fairness dimensions identified concrete technical deficiencies that required improvements. The baseline model demonstrated inadequate demographic representation, insufficient fairness evaluation data and low prediction confidence levels not apparent from the original accuracy metrics.

These findings directly guided the system enhancements we discussed. Dataset improvements addressed demographic representation gaps, architectural improvements reduced prediction uncertainty and we conducted a fairness analysis \citep{poretschkin_ai_2023}. The enhanced system achieves high accuracy with good demographic fairness and robustness across augmentation types, representing substantial improvement over the baseline model.

This improvement cycle demonstrates the catalogue's primary value: translating abstract trustworthiness principles into concrete evaluation criteria that drive technical enhancement \citep{poretschkin_ai_2023}. Organizations using structured assessment frameworks during development can therefore create systems better positioned for eventual regulatory compliance, even though completing the assessment does not mean achieving compliance with the AI Act \citep{fpf_conformity-assessments_2025}.

\subsection{Remaining Requirements for Compliance}
\label{sec:remaining-requirements}

Full AI Act compliance for emotion recognition systems requires additional work beyond technical assessment alone. First, organizations must analyze whether the system falls under prohibited use under Article 5. Emotion recognition in workplace and educational contexts became prohibited on February 2, 2025, except for medical or safety purposes \citep{european_commission_ai_2024, turner_eu_2025}.
For permissible deployments, organizations must establish quality management systems meeting Article 17 requirements, prepare technical documentation, implement post-market monitoring systems and conduct formal conformity assessment before the August 2, 2026 deadline for high-risk systems \citep{fpf_conformity-assessments_2025, european-union_regulation_2024}. These requirements extend beyond technical system characteristics into organizational processes and ongoing compliance commitments \citep{modulos_eu_2025}.

However, systems developed using structured assessment frameworks like the Fraunhofer Catalogue can benefit from systematic documentation, identified improvement areas and demonstrated attention to trustworthiness dimensions. This preparatory work, while insufficient for compliance alone, can ease the path toward formal conformity assessment when harmonized standards become available and organizational compliance infrastructure develops \citep{a-lign_understanding_2025}. These findings position the Fraunhofer framework as valuable for development guidance while highlighting the substantial work remaining for formal compliance.

\section{Findings}
\label{sec7}
This section addresses our three research questions, using the insights we gained from the entire process of technical analysis, system enhancements and recertification with the Fraunhofer Catalogue.

\subsection{RQ1: Practical Insights from Complete Self-Certification}
\label{sec:rq1}

\textit{\textbf{RQ1:} What practical insights emerge from conducting a complete self-certification cycle of an AI system using the Fraunhofer AI Assessment Catalogue?}

\subsubsection{Certification Drives Deep Technical Investigation}

The Fraunhofer Catalogue's structured criteria necessitated deeper technical investigation and reengineering to fulfill its requirements. Basic accuracy documentation obscured the reliability issue of low confidence scores in the predictions. The fairness dimension required demographic metadata absent from the original datasets, which led us to use a more complete dataset.
This creates implications for legacy systems: systems that cannot be modified must satisfy every requirement in their existing state. The baseline EmoPy system exemplified this limitation, necessitating complete reconstruction to allow for the necessary improvements.

\subsubsection{Catalogue Functions as Development Framework}

Beyond its certification purpose, the catalogue proved valuable as a development tool. Rather than serving only as a final assessment, the structured questions across trustworthiness dimensions provided effective guidance for system refinements.
This directly guided our enhancements, translating abstract principles into measurable performance and an overall better system. Organizations can leverage this proactively during development to create better AI systems.

\subsubsection{Documentation Emerges from Development Integration}

Integrating this type of certification process throughout development created documentation as a natural byproduct. Addressing catalogue criteria during implementation ensured that documentation captured actual system properties, strengthening the validity of the certification.

\subsection{RQ2: Technical Testing versus Documentation-Only Certification}
\label{sec:rq2}

\textit{\textbf{RQ2:} What does technical testing reveal that documentation-only certification can miss?}

\subsubsection{Technical Testing Can Reveal Issues}

Technical implementation exposed problems that we did not find previously in documentation. The baseline assessment showed acceptable documented accuracy that masked unreliability: high entropy and low confidence. Basic performance documentation focusing on accuracy alone missed these characteristics.
Our enhanced model evaluation used more comprehensive metrics: entropy, confidence, demographic fairness and augmentation robustness. These metrics document important system behavior for a fair and certifiable system.

\subsubsection{Complementary Roles, Not Replacement}

Technical testing and documentation serve complementary functions. The Fraunhofer AI Assessment Catalogue relies primarily on documentation of the system. However, examining actual technical details more deeply can lead to a more complete picture and help avoid oversights.

\subsection{RQ3: Relationship to AI Act Requirements}
\label{sec:rq3}

\textit{\textbf{RQ3:} To what extent does this self-certification address the requirements established by the AI Act?}

\subsubsection{Fundamental Gaps Prevent Direct Compliance}

Harmonized European standards remain under development and the Fraunhofer Catalogue does not constitute a harmonized standard. Without published standards, systems cannot claim presumption of conformity regardless of the assessment quality.

\subsubsection{Substantial Preparatory Value Despite Gaps}

Despite these gaps, structured self-certification provides substantial preparatory value. The systematic improvements and documentation we created through this process align with AI Act principles and can serve as a foundation for formal conformity assessment when harmonized standards become available.

%%%%%%%%%%%%%%%%%%%%%%%%%%%%%%%
\section{Conclusion and Future Research Directions}
\label{sec8}

Our research into the self-certification of the facial emotion recognition AI system yielded insights regarding the Fraunhofer certification framework, system development and regulatory compliance. The complete self-certification cycle using the Fraunhofer AI Assessment Catalogue demonstrated its effectiveness as both an evaluation tool and development framework. The enhanced emotion recognition system satisfies the certification requirements for the reliability and fairness dimensions we examined, successfully addressing deficiencies we identified during initial assessment through systematic technical improvements.

Integrating catalogue requirements into the development process led to concrete improvements: enhanced architecture reduced prediction uncertainty, expanded datasets enabled demographic fairness analysis and systematic documentation emerged naturally from the process. This demonstrates that the certification framework functions as a proactive development guide rather than an evaluation tool.

However, substantial gaps separate structured self-certification from AI Act compliance. The absence of a completed set of harmonized European standards prevents systems from claiming presumption of conformity regardless of assessment quality and the Fraunhofer Catalogue does not provide the mandatory conformity assessment procedures required for compliance with the AI Act. This underscores the critical distinction between internal quality assessment and legal compliance demonstration.

Our research addressed its core objectives by providing practical insights into a complete self-certification process, demonstrating how technical implementation reveals issues that documentation alone could miss and demonstrating the preparatory value of structured assessment despite compliance gaps. Future work could investigate how certification frameworks evolve once harmonized standards become available and explore how organizations can effectively transition from self-certification to formal conformity assessment when regulatory requirements are fully available. Additionally, attempting a certification of a real-world AI model deployed by a company would yield valuable insights into the practical challenges of the certification process and the steps required to achieve AI Act compliance in commercial contexts.

\vspace{2mm} \noindent \textbf{Acknowledgements:} This work was supported by the FFG COMET program and the strategic COMET project ``KnowCERTIFAI'' led by Know Center. 

\bibliography{biblography_main_man}

\end{document}